\documentclass{article}
\usepackage{amsmath}
\usepackage[margin=1in]{geometry}
\usepackage{amsfonts}
\usepackage{multirow}
\usepackage{graphicx}
\usepackage{booktabs}
\usepackage{xcolor}
\usepackage{hyperref}
\usepackage{graphicx,verbatim}
\begin{document}
\title{Efficient Conformal Volumetry for Template-Based Segmentation}
%

\author{Matt Y. Cheung, Ashok Veeraraghavan \& Guha Balakrishnan 
\\
Department of Electrical \& Computer Engineering\\
Rice University \\
Houston, TX 77005, USA \\
}
  
\maketitle              
\begin{abstract}
Template-based segmentation, a widely used paradigm in medical imaging, propagates anatomical labels via deformable registration from a labeled atlas to a target image, and is often used to compute volumetric biomarkers for downstream decision-making. 
While conformal prediction (CP) provides finite-sample valid intervals for scalar metrics, existing segmentation-based uncertainty quantification (UQ) approaches either rely on learned model features, often unavailable in classic template-based pipelines, or treat the registration process as a black box, resulting in overly conservative intervals when applied directly in output space.
We introduce ConVOLT, a CP framework that achieves efficient volumetric UQ by conditioning calibration on properties of the estimated deformation field from template-based segmentation.
ConVOLT calibrates a learned volumetric scaling factor from deformation space features.
We evaluate ConVOLT on template-based segmentation tasks involving global, regional, and label volumetry across multiple datasets and registration methods. ConVOLT achieves target coverage while producing substantially tighter intervals than output-space conformal baselines.
Our work paves way to exploit the registration process for efficient UQ in medical imaging pipelines.
Code available at \url{https://github.com/matthewyccheung/convolt}.
\end{abstract}
\section{Introduction}
Obtaining statistical guarantees on clinically meaningful quantities derived from medical image segmentation maps, such as anatomical volumes, is crucial to support reliable downstream decision-making. 
Recent work has demonstrated that Conformal Prediction (CP)~\cite{vovk2005algorithmic,fontana2023conformal,shafer2008tutorial} can provide finite-sample valid coverage guarantees for scalar metrics derived from black-box imaging pipelines~\cite{cheung2025probabilistic}. 
But while this black-box approach guarantees marginal coverage, it generally yields inefficient (wide) prediction intervals. Recent work has shown that we can produce far more efficient intervals for metrics derived from medical image segmentation deep networks by conditioning on their internal learned feature representations (``feature CP'')~\cite{cheung2025compass,teng2022predictive}. 
However, many clinical pipelines rely on a different segmentation paradigm called \emph{template-based segmentation}~\cite{iglesias2015multi,cabezas2011review}, where a new image is segmented by densely registering it to a labeled template and propagating the template's labels through the deformation field~\cite{sotiras2013deformable,klein2009evaluation,balakrishnan2019voxelmorph}. 
Because this paradigm does not use an explicit segmentation model, it is unclear how to readily extend feature CP methods to obtain efficient prediction intervals. 

\begin{figure}[t]
    \centering
    \includegraphics[width=\linewidth]{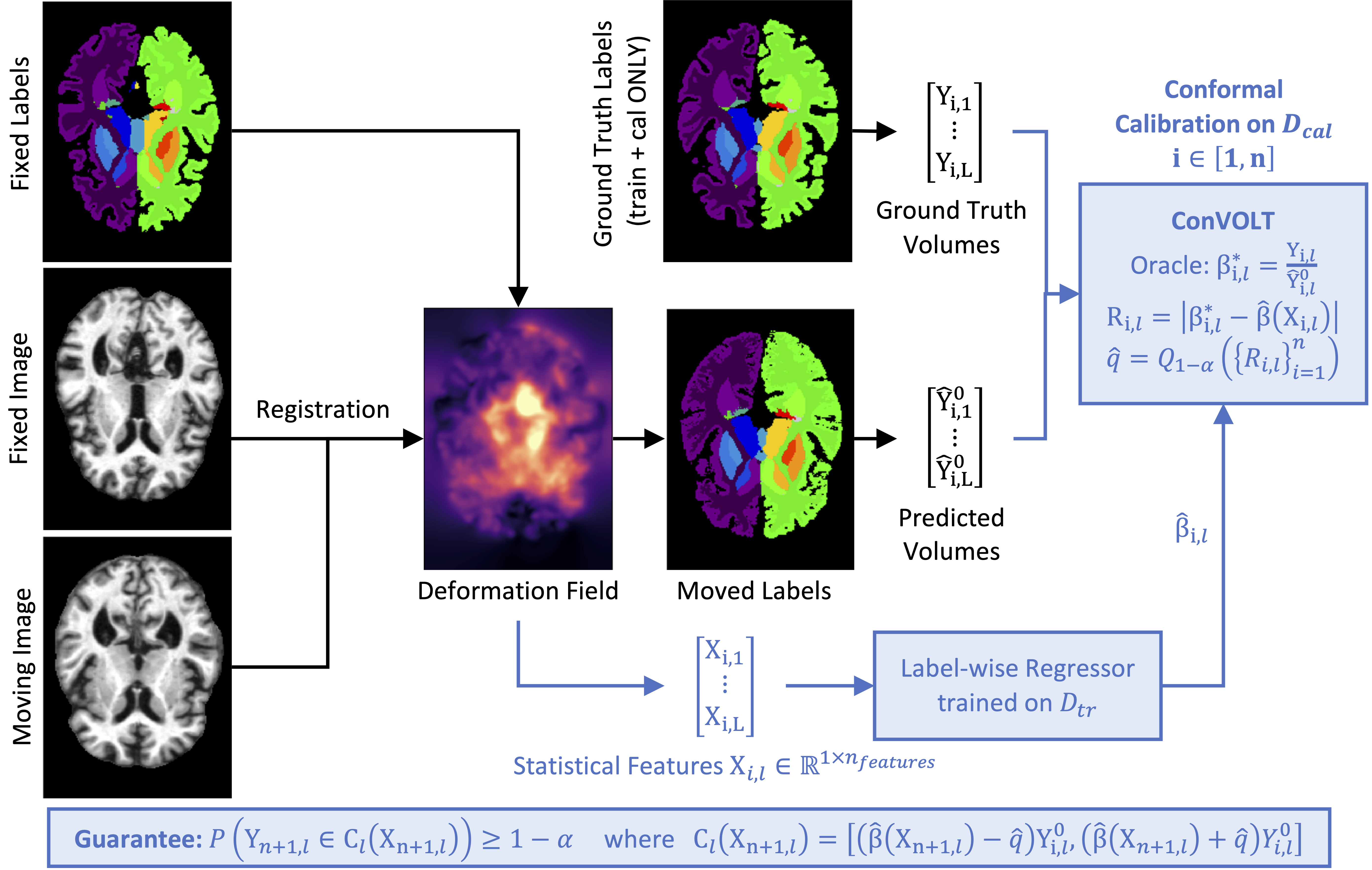}
    \caption{\textbf{Overview of ConVOLT.}
    We show a standard template-based segmentation pipeline for case $i$ (black) where a moving image is registered to a fixed image, resulting in a deformation field.
    The fixed labels are warped to compute predicted labels and volumes $\widehat{Y}^0_{i,l}$ on the moved image.
    ConVOLT (blue) trains a label-specific regressor on the training dataset $D_{tr}$ with statistical features extracted from the deformation field features to predict a multiplicative scaling factor $\widehat{\beta}_{i,l}$. 
    The scaling factor is conformalized using the calibration dataset $D_{cal}$ to attain valid prediction intervals. 
    }
    \label{fig:overview}
\end{figure}

In this work, we bridge this gap by showing that conditioning CP on properties of the \emph{deformation field} can yield substantially tighter intervals while preserving coverage. 
We hypothesize that deformation-aware conditioning leads to improved efficiency over output-space CP baselines because many downstream metrics are deterministic functionals of the deformation. 
In fact, volumetric change can be written as an integral of the Jacobian determinant over spatial regions, so uncertainty in the biomarker is induced by uncertainty in the deformation itself. 
The deformation field encodes informative geometric cues -- such as local expansion, contraction, and spatial heterogeneity -- that govern variability in volume estimates, with highly heterogeneous Jacobians typically inducing greater volumetric uncertainty than smooth, near-identity transformations.

Based on these observations, we introduce \textbf{ConVOLT} (Conformal Volumetry for Template-Based Segmentation) (Fig.~\ref{fig:overview}), a CP framework for template-based segmentation pipelines that calibrates a learned volumetric scaling factor from deformation space features.
The scaling preserves non-negativity and scales uncertainty with volume.
This yields finite-sample valid prediction intervals for volumetric biomarkers while exploiting geometric structure to improve efficiency. 
We evaluate ConVOLT on template-based segmentation tasks involving global, regional, and label volumetry across multiple datasets and registration methods, and find that ConVOLT achieves target coverage while producing tighter intervals than output-space CP baselines. 
These results demonstrate that exploiting deformation structure can significantly improve the efficiency of uncertainty quantification for template-based segmentation pipelines.

\section{Method}
\label{sec:convolt}

\noindent The key idea in ConVOLT is to calibrate \emph{multiplicative} deviations induced by the deformation field rather than additive residuals in output space. 
Because many volumetric biomarkers are deterministic functionals of the deformation (e.g., integrals of the Jacobian determinant), multiplicative errors in the deformation propagate multiplicatively to the scalar target. 
ConVOLT exploits this structure by learning deformation-conditioned scaling factors and conformalizing them via split CP, yielding closed-form volumetric intervals with finite-sample coverage.
\newline

\noindent\textbf{Problem setup.}
For each case $i\in\{1,\dots,N\}$, a registration algorithm produces a displacement field $u_i:\Omega\to\mathbb{R}^3$ on a voxel lattice $\Omega$. 
The induced local volume change is given by the Jacobian determinant $J_i(x)=\det(\mathbf I+\nabla u_i(x))$. 
Let $m_{i,l}^{F}:\Omega\to\{0,1\}$ denote a fixed-space mask for label $l\in\{1,\dots,L\}$, where labels may represent the global ROI, spatial regions (e.g., shells), or anatomical labels. 
All quantities are evaluated on the fixed grid.
Let $dV$ denote voxel volume. 
Integrating the Jacobian over a label mask yields the deformation-derived baseline prediction
$\widehat Y^0_{i,l}=\sum_{x\in\Omega} m_{i,l}^{F}(x)\,J_i(x)\,dV$ with ground-truth scalar $Y_{i,l}$. 
\newline

\noindent\textbf{Split CP. } We seek valid prediction intervals for $Y_{i,l}$ that adapt to deformation variability. 
We utilize the following lemma to prove validity: 
Let $(X_{i,l},Y_{i,l})$ be exchangeable pairs and split the data into a training set $D_{\mathrm{tr}}$ and calibration set $D_{\mathrm{cal}}=\{(X_{i,l},Y_{i,l})\}_{i=1}^n$. 
Given non-conformity scores $R_{i,l}$ on $D_{\mathrm{cal}}$, define $\widehat q$ as the $\lceil(1-\alpha)(n+1)\rceil$-th order statistic. 
For a test pair $(X_{\text{test},l},Y_{\text{test},l})$, the conformal set $\{y:\ R(X_{\text{test},l},y)\le \widehat q\}$ satisfies $P(Y_{\text{test},l}\in S_{\widehat q}(X_{\text{test},l}))\ge 1-\alpha$ under exchangeability.
This is a standard result of split CP~\cite{vovk2005algorithmic,shafer2008tutorial,fontana2023conformal}.
The remainder of this section focuses on how to construct a novel volumetric non-conformity score and its valid prediction interval that leverage deformation field features from template-based segmentation.
\newline

\noindent\textbf{ConVOLT. }
We start with the observation that multiplicative perturbations in the deformation induce multiplicative deviations in $\widehat Y^0_{i,l}$.
This is because volumetric targets are integrals of $J_i(x)$.
We model this via $\widehat Y^\beta_{i,l}=\beta_{i,l}\widehat Y^0_{i,l}$ where $\beta_{i,l}\ge0,$ and define the oracle ratio $\beta^\star_{i,l}=Y_{i,l}/\widehat Y^0_{i,l}$. 
This parameterization preserves non-negativity and scales uncertainty with volume.
Next, using the training set $D_{tr}$, we learn a predictor for deformation-conditioned ratios $\widehat\beta(X_{i,l})$ from deformation-derived features $X_{i,l}$ (e.g., Jacobian statistics, spatial heterogeneity) to $\beta^\star_{i,l}$ on the training dataset $D_{tr}$.
On $D_{\mathrm{cal}}$, define nonconformity scores $R_{i,l}=|\beta^\star_{i,l}-\widehat\beta(X_{i,l})|$ and compute $\widehat q$ by taking the $\lceil(1-\alpha)(n+1)\rceil$-th empirical quantile of the scores.
For a test case with features $x$ and baseline $\widehat Y^0_l(x)$, the conformal set in ratio space $\{y:\ |y/\widehat Y^0_l(x)-\widehat\beta(x)|\le\widehat q\}$ yields the interval:
\begin{equation}
    \label{eq:convoltset}
    \mathcal C_l(x)=\big[(\widehat\beta(x)-\widehat q)\widehat Y^0_l(x),\,(\widehat\beta(x)+\widehat q)\widehat Y^0_l(x)\big].
\end{equation}
\noindent The validity of ConVOLT relies on the following proposition: 
Assume $(X_{i,l},Y_{i,l})$ are exchangeable across calibration and test cases. 
Then Eq.~\ref{eq:convoltset} satisfies $P(Y_{n+1,l}\in\mathcal C_l(X_{n+1,l}))\ge 1-\alpha$.
The result follows from the split CP lemma, $P(|\beta^\star_{n+1,l}-\widehat\beta(X_{n+1,l})|\le\widehat q)\ge 1-\alpha$. 
Since $Y_{n+1,l}=\beta^\star_{n+1,l}\widehat Y^0_{n+1,l}$ and the mapping $\beta\mapsto Y$ is monotone for $\beta\ge0$, the event transfers directly to $\mathcal C_l$.
\newline

\noindent When multiple labels share a deformation field, we aggregate label-wise scores into a case-level score $S_i=A(\{R_{i,l}\}_{l=1}^L)$ where $A:\mathbb{R}^L\rightarrow\mathbb{R}$ is an aggregation function (e.g., $\max_l$ for simultaneous guarantees or quantile for weaker guarantees) and apply split CP to $\{S_i\}_{i\in D_{\mathrm{cal}}}$. 
The resulting $\widehat q$ is shared across labels, yielding case-level coverage under case-level exchangeability.
\newline

\section{Experimental Setup}
\label{sec:experiments}
\noindent\textbf{Datasets and Volume Metrics. }
We evaluated ConVOLT on intra-patient and inter-patient template-based segmentation tasks using datasets from the Learn2Reg challenge~\cite{hering2022learn2reg}: 1) NLST: 209 lung CT inhale-exhale pairs with lung masks, 2) ThoraxCBCT: 30 chest CT pairs across breathing phases with lung masks, and 3) OASIS: 409 brain MRI scans with 35 anatomical labels.
In the intra-patient setting (ThoraxCBCT and NLST), each example is a longitudinal image pair and the target is volume change. 
In the inter-patient setting (OASIS), atlas images are registered to a target patient and warped labels yield absolute-volume targets. 
To prevent data leakage, we excluded atlases used for training from calibration and testing sets.
We evaluate global volume/volume change (all datasets), regional volume change (ThoraxCBCT and NLST), and label volume (OASIS).
For regional volume change, we partitioned each lung mask into $L=5$ concentric radial shells from boundary to interior. 
We defined each shell as label $l$ with target $Y_{i,l}$ equal to regional volume change. 
We evaluated patient-level guarantees by aggregating label-wise scores using $\max$ and $Q_{0.9}(\cdot)$ aggregation. We performed calibration at the patient level.
For label volume, we define each of the 35 anatomical structures as label $l$ with target absolute volume $Y_{i,l}$ and atlas-derived prediction $\widehat Y^0_{i,l}$. 
We extracted deformation-conditioned features near predicted label boundaries and fit a separate ridge regressor per label to predict multiplicative ratios. 
We applied calibration independently per label.
\newline

\noindent\textbf{Registration Algorithms.}
We used two popular, representative registration methods: Demons~\cite{thirion1996non} and VoxelMorph~\cite{balakrishnan2019voxelmorph}. 
We used a standard multi-resolution configuration for Demons.
We trained VoxelMorph with normalized cross-correlation similarity and smoothness regularization. We kept all preprocessing and training settings consistent across methods and datasets.
\newline

\noindent\textbf{Experimental Protocol and Baselines. }
For each dataset, we randomly split the patient IDs into training/calibration/test sets with 0.4/0.4/0.2 proportions for NLST and OASIS, and 10/15/5 for ThoraxCBCT.
We repeated experiments 100 times with random calibration–test splits at miscoverage level $\alpha=0.1$. 
We report mean and standard deviation of empirical coverage and interval size. For global volume/volume change, we compared the following methods:
\begin{itemize}
\item \textbf{SCP}~\cite{lei2018distribution}: The absolute residual score $R_{i,l}=|Y_{i,l}-\widehat Y^0_{i,l}|$.
\item \textbf{CQR}~\cite{romano2019conformalized}: A conditional quantile regression with pinball loss and ridge regularization ($\lambda=0.01$).
\item \textbf{LCP}~\cite{papadopoulos2002inductive}: A locally calibrated residual CP using $k=50$ nearest calibration neighbors based on $|\widehat Y^0|$.
\item \textbf{ConVOLT}: Our proposed deformation-space method using ridge regression to predict multiplicative ratios. We used features: $\log J$ statistics (mean, std, mean absolute), Jacobian quantiles (10th, 50th, 90th percentile), folding fractions ($P(J<0.1)$, $P(J<0.01)$), displacement magnitude (mean, 90th percentile, max), $\|\nabla\log J\|$ statistics (mean, 90th percentile, max), divergence/curl summaries (mean, 90th percentile, max), and fixed--warped similarity residuals (mean absolute error, mean squared error, correlation).
\end{itemize}
\begin{table}[t]
\caption{\textbf{ConVOLT achieves high efficiency compared to baselines for global volume guarantees.} 
For $\alpha=0.1$, we show interval size and coverage (Mean$\pm$STD) for SCP, CQR, LCP, and ConVOLT. 
While ConVOLT performs well on the majority of benchmarks, it achieves slightly worse/comparable performance on NLST with Demons and ThoraxCBCT with VoxelMorph.}
\centering
{\fontsize{8}{10}\selectfont
{%
\begin{tabular}{@{}cccccc@{}}

\toprule
 &  & \multicolumn{2}{c}{Demons} & \multicolumn{2}{c}{VoxelMorph} \\ \midrule
Dataset & Method & Coverage & Interval Size (mL) & Coverage & Interval Size (mL) \\ \midrule
\multirow{4}{*}{ThoraxCBCT} & CQR & 0.90±0.14 & 2707±951 & 0.91±0.15 & 2355±830 \\
 & ConVOLT & 0.93±0.12 & \textbf{1222±357} & 0.93±0.14 & \textbf{2354±1018} \\
 & LCP & 0.91±0.13 & 4631±986 & 0.91±0.15 & 2400±542 \\
 & SCP & 0.91±0.13 & 4631±986 & 0.91±0.15 & 2400±542 \\ \midrule
\multirow{4}{*}{NLST} & CQR & 0.91±0.06 & \textbf{496±84} & 0.91±0.06 & 446±117 \\
 & ConVOLT & 0.91±0.06 & 511±89 & 0.92±0.05 & \textbf{405±86} \\
 & LCP & 0.90±0.06 & 685±77 & 0.94±0.05 & 736±65 \\
 & SCP & 0.91±0.06 & 895±121 & 0.91±0.06 & 721±56 \\ \midrule
\multirow{4}{*}{OASIS} & CQR & 0.90±0.04 & 130±6 & 0.90±0.05 & 129±6 \\
 & ConVOLT & 0.91±0.04 & \textbf{105±7} & 0.91±0.06 & \textbf{106±11} \\
 & LCP & 0.90±0.04 & 152±8 & 0.89±0.05 & 134±5 \\
 & SCP & 0.91±0.03 & 159±6 & 0.91±0.05 & 138±5 \\ \bottomrule
\end{tabular}%
}
}
\label{tab:totalvol}
\end{table}

\section{Results and Discussion}

\begin{figure}[t]
    \centering
    \includegraphics[width=\linewidth]{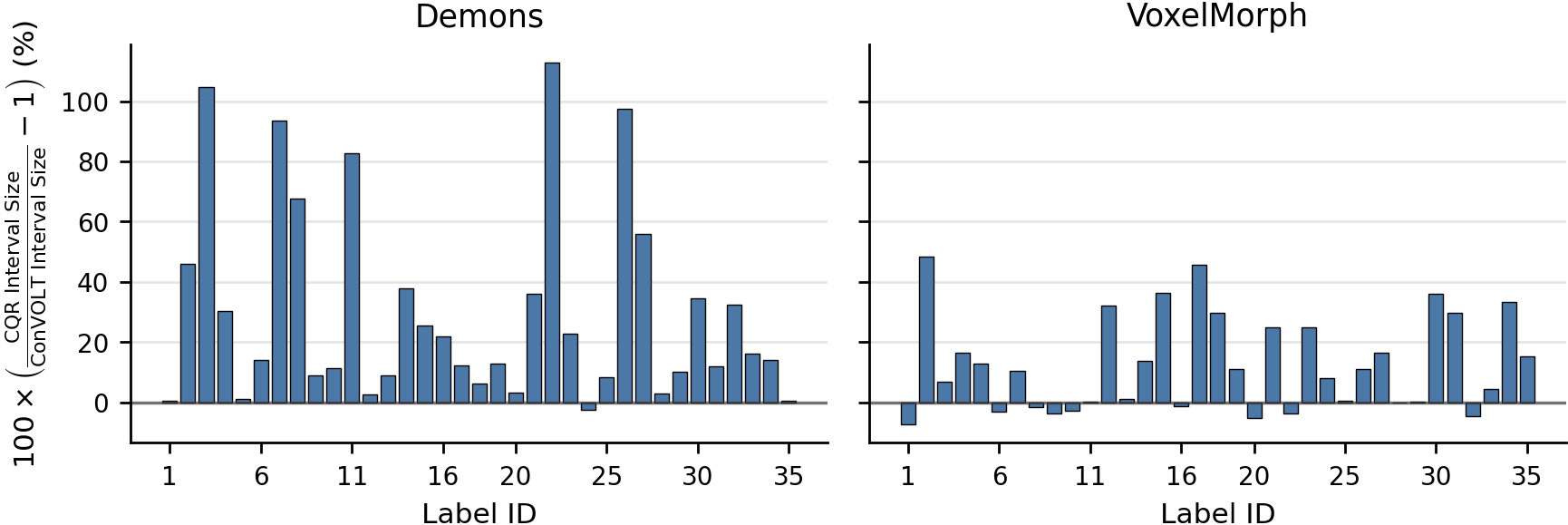}
    \caption{\textbf{ConVOLT achieves high efficiency compared to CQR for label volume guarantees. } For $\alpha=0.1$, we show interval size inflation of CQR (best performing baseline) in \% for each label in OASIS with ConVOLT as the baseline. We find that ConVOLT achieves higher efficiency on the majority of labels.}
    \label{fig:labels}
\end{figure}

\noindent\textbf{ConVOLT produces efficient CP intervals. } 
For total volume change (ThoraxCBCT and NLST) and volume (OASIS), we compared ConVOLT to baseline methods in Tab.~\ref{tab:totalvol}. 
Across most datasets and registration algorithms, ConVOLT achieved the desired coverage while producing smaller or comparable intervals relative to baselines. 
These results suggest that conditioning on deformation-field features can yield efficiency gains beyond those obtained from adaptive output-space scores alone. 
However, in a few settings (e.g., NLST with Demons and ThoraxCBCT with VoxelMorph), ConVOLT and CQR achieved similar interval sizes. 
In these cases, residual volumetric variability is largely explained by volume magnitude itself, and the additional deformation features provided limited predictive signal for correcting errors.
We also compared interval size for regional volume change in ThoraxCBCT and NLST against CQR in Tab.~\ref{tab:regional} for aggregation functions $\max$ and $Q_{0.9}$. 
On ThoraxCBCT, ConVOLT consistently produced tighter intervals while maintaining coverage, indicating that deformation features captured informative structure for regional errors. 
On NLST, results were mixed: ConVOLT improved efficiency for VoxelMorph but was less efficient than CQR for Demons under the $\max$ aggregation. 
This suggests that for NLST with Demons, the deformation-derived features were weakly correlated with regional registration errors, so conditioning on them introduced additional variance without improving bias correction. 
Finally, we compared interval size for label volume in OASIS against CQR in Fig.~\ref{fig:labels}.
Our results indicated that ConVOLT produced tighter intervals while maintaining coverage for the majority of labels.
Overall, these findings indicate that deformation-aware calibration is most beneficial when deformation features meaningfully explain volumetric error variability, and remains competitive with strong adaptive baselines otherwise.
\newline

\begin{table}[t]
\caption{\textbf{ConVOLT achieves high efficiency compared to CQR for regional guarantees. } 
For $\alpha=0.1$, we show interval size and coverage (Mean$\pm$STD) for CQR (best performing baseline) and ConVOLT.
While ConVOLT performs well on the majority of benchmarks, it has noticeably worse performance on NLST.}
\centering
{\fontsize{8}{10}\selectfont
{%
\begin{tabular}{@{}ccccccc@{}}
\toprule
 &  &  & \multicolumn{2}{c}{Demons} & \multicolumn{2}{c}{VoxelMorph} \\ \midrule
Dataset & Agg & Method & Coverage & Interval Size (mL) & Coverage & Interval Size (mL) \\ \midrule
\multirow{4}{*}{ThoraxCBCT} & \multirow{2}{*}{max} & CQR & 0.91±0.14 & 1381±362 & 0.92±0.13 & 1148±273 \\
 &  & ConVOLT & 0.92±0.14 & \textbf{1117±146} & 0.95±0.10 & \textbf{1128±295} \\
 & \multirow{2}{*}{$Q_{0.9}$} & CQR & 0.91±0.14 & 1139±187 & 0.92±0.14 & 898±154 \\
 &  & ConVOLT & 0.92±0.14 & \textbf{743±111} & 0.95±0.11 & \textbf{822±176} \\ \midrule
\multirow{4}{*}{NLST} & \multirow{2}{*}{max} & CQR & 0.91±0.05 & \textbf{670±39} & 0.91±0.05 & 376±34 \\
 &  & ConVOLT & 0.91±0.05 & 1083±91 & 0.91±0.05 & \textbf{357±40} \\
 & \multirow{2}{*}{$Q_{0.9}$} & CQR & 0.91±0.05 & \textbf{619±28} & 0.92±0.05 & 338±30 \\
 &  & ConVOLT & 0.91±0.05 & 744±58 & 0.91±0.05 & \textbf{273±29} \\ \bottomrule
\end{tabular}%
}
}
\label{tab:regional}
\end{table}

\begin{figure}[t]
    \centering
    \includegraphics[width=\linewidth]{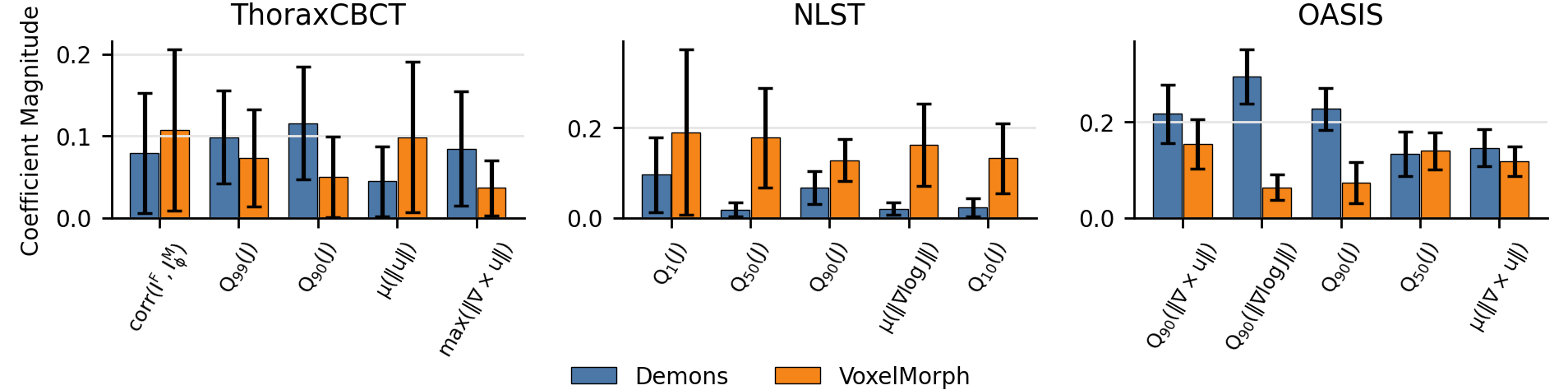}
    \caption{\textbf{Coefficient magnitudes explain ConVOLT's efficiency improvements. } 
    We show Mean$\pm$STD of absolute ridge coefficients for features across 100 random calibration-test splits. 
    Large and stable magnitudes indicate features are consistently used to predict multiplicative correction, while small and unstable magnitudes indicate the limited predictive value of features and explains weaker performance.
    }
    \label{fig:coefstability}
\end{figure}

\noindent\textbf{Interpretability. }
ConVOLT learns multiplicative correction factors $\widehat{\beta}_{i,l}$ using ridge regression on deformation-derived features, allowing direct interpretation through the learned regression coefficients. 
We plotted the top-5 coefficient magnitudes over 100 randomized patient-level train/calibration/test splits and reported the mean and standard deviation of the coefficient magnitude for each feature in Fig.~\ref{fig:coefstability}. 
Larger magnitudes indicate higher contribution.
On ThoraxCBCT, multiple deformation features exhibited moderate magnitudes for both Demons and VoxelMorph, suggesting that volumetric errors are partially explained by deformation heterogeneity. 
This aligns with the observed efficiency gains of ConVOLT on this dataset. 
On OASIS, Jacobian-based quantile features dominated with consistently large coefficients, indicating that spatial expansion statistics are strong predictors of volumetric bias.
Correspondingly, ConVOLT achieved substantial interval reductions. 
In contrast, for NLST with Demons, coefficient magnitudes were relatively small, suggesting that deformation-derived features provide limited explanatory power for volumetric error. 
This explains why ConVOLT offers little or no efficiency improvement in this setting: when features are weakly correlated with residual error, deformation-aware conditioning cannot meaningfully outperform adaptive output-space baselines.
\newline

\begin{table}[t]
\caption{\textbf{Multiplicative modeling, learned ratios, and localized features improve ConVOLT efficiency. }
We report interval inflation (\%) relative to ConVOLT when replacing multiplicative modeling with additive corrections, removing learning or features, or using global instead of local deformation features.}
\centering
{\fontsize{8}{10}\selectfont
{%
\begin{tabular}{@{}cccccc@{}}
\toprule
 &  & \multicolumn{2}{c}{Demons} & \multicolumn{2}{c}{VoxelMorph} \\ \midrule
Dataset & Ablation & Coverage & Inflation (\%) & Coverage & Inflation (\%) \\ \midrule
\multirow{4}{*}{ThoraxCBCT} & Additive & 0.93±0.13 & 41.4±56.1 & 0.94±0.14 & 2.8±60.4 \\
 & No Features & 0.91±0.14 & 93.4±77.3 & 0.91±0.15 & -22.2±38.5 \\
 & No Learning & 0.91±0.13 & 218.1±108.4 & 0.90±0.14 & -7.1±45.5 \\
 & Global Features & 0.92±0.14 & 73.8±32.9 & 0.95±0.11 & 65.7±57.8 \\ \midrule
\multirow{4}{*}{NLST} & Additive & 0.91±0.06 & 3.3±26.3 & 0.91±0.06 & 0.2±28.5 \\
 & No Features & 0.91±0.06 & 11.9±27.9 & 0.92±0.05 & 12.8±26.0 \\
 & No Learning & 0.92±0.06 & 75.1±36.2 & 0.91±0.06 & 64.0±36.6 \\
 & Global Features & 0.91±0.05 & 31.7±11.5 & 0.91±0.05 & 72.2±21.3 \\ \midrule
\multirow{3}{*}{OASIS} & Additive & 0.90±0.04 & 0.0±9.4 & 0.90±0.06 & -0.9±14.0 \\
 & No Features & 0.90±0.03 & 27.6±9.7 & 0.90±0.05 & 26.4±16.2 \\
 & No Learning & 0.91±0.03 & 53.3±11.7 & 0.90±0.05 & 30.2±14.3 \\ \bottomrule
\end{tabular}%
}
}
\label{tab:ablation}
\end{table}

\noindent\textbf{Ablations.} 
ConVOLT's efficiency can be attributed to two factors: 1) learned scores from deformation field features and 2) adaptive scores using the multiplicative formulation. 
We made several intentional design choices for ConVOLT. 
To justify them, we performed ablations on 1) multiplicative ($\beta_i \widehat{Y}_{0,i}$) versus additive scalars ($\widehat{Y}_{0,i}+\beta_i$), 2) no learning ($\hat{\beta}(x)=1$), 3) no features ($\hat{\beta}(x)=\mu(\hat{\beta}_i^*)$), and 4) localized features (region-restricted deformation features) versus global features (whole-mask deformation features) for regional guarantees.
We show set size inflation relative to ConVOLT in Tab.~\ref{tab:ablation}.
We found that multiplicative factors produced more efficient intervals than additive scalars, learning and sample-specific features improved efficiency, and localized features improved efficiency for region based guarantees.
\newline

\section{Conclusion}
We demonstrated with ConVOLT that conditioning CP on deformation-field features can produce substantially more efficient CP intervals for deformation-derived volumetric biomarkers. 
Our results further showed that its effectiveness depends on the predictive strength of geometric features. 
More broadly, these findings suggest that exploiting structure within medical imaging pipelines, beyond black-box outputs, can enable more efficient and interpretable UQ for clinically-relevant downstream metrics.

\subsubsection*{Acknowledgements}
MC would like to acknowledge support from a fellowship from the Gulf Coast Consortia on the NLM Training Program in Biomedical Informatics and Data Science T15LM007093.
\bibliographystyle{splncs04}
\bibliography{ref}
\end{document}